\documentclass[fdp,a4paper,fleqn]{w-art}
\usepackage{times,cite,w-thm}
\usepackage[english]{babel}
\usepackage[]{graphicx}
\usepackage{graphics}
\usepackage{amssymb}
\usepackage{epsf}
\usepackage{rotating}
\usepackage{axodraw}
\usepackage{pstricks}
\usepackage{slashed}
\newcommand{\bea}{\begin{eqnarray}}
\newcommand{\eea}{\end{eqnarray}}

\def\beq{\begin{equation}}
\def\eeq{\end{equation}}

\def\beqn{\begin{eqnarray}}
\def\eeqn{\end{eqnarray}}
\def\beqa{\begin{eqnarray}}
\def\eeqa{\end{eqnarray}}
\def\ba{\begin{eqnarray}}
\def\ea{\end{eqnarray}}
\begin{document}
\pagespan{1}{}


\title[Effective Actions with Gauge and Conformal Anomalies]{The Effective Actions of Pseudoscalar and Scalar Particles\\ in Theories with Gauge and Conformal Anomalies {\footnote{Presented by Claudio Corian\`{o} at the 9th Hellenic Workshop on Elementary Particle Physics and Gravity, Corfu Summer Institute, Greece August 30 September 20, 2009}}}

\author[R. Armillis]{Roberta Armillis\inst{1}}%
\address[\inst{1}]{Dipartimento di Fisica, Universit\`{a} del Salento \\
and  INFN Sezione di Lecce, Via Arnesano 73100 Lecce, Italy}
\author[C. Corian\`{o}]{Claudio Corian\`{o}\inst{1}%
 \footnote{Corresponding author\quad E-mail:~\textsf{claudio.coriano@le.infn.it},
          Phone: +39\,0832\,297\,437.
            }}
\author[L. Delle Rose]{Luigi Delle Rose\inst{1}}%
\author[M. Guzzi]{Marco Guzzi\inst{2}}
\address[\inst{2}]{Department of Physics, Southern Methodist University, \\Dallas TX 75275, USA\\}
\author[A. Mariano]{Antonio Mariano\inst{1}}%
\begin{abstract}
We review recent work on the effective field theory description and the phenomenology of axion-like and scalar particles in models characterized by gauge and/or conformal anomalies.
\end{abstract}
\maketitle

 \renewcommand{\leftmark}
{R. Armillis et al.: Effective Actions with Gauge and Conformal Anomalies}

\section{Introduction}

Massless moduli fields, in the form of scalars and pseudoscalars, are intrinsic components of effective theories below the Planck scale, possibly derived from a string theory, which acquire some mass as the universe expands and its temperature decreases.  They share features which are quite similar to those of Nambu-Goldstone modes of broken gauge symmetries.

Moduli have no potential, except for those triggered by possible non-perturbative effects encountered in the early universe, for instance at the electroweak or at the QCD phase transitions. Historically, the first example of such a case is the breaking of the global Peccei-Quinn (PQ) symmetry, which leaves an axion as a dark matter candidate, with all its associated topological structures, in the form of cosmic strings and domain walls, which are expected to be washed out by inflation if the breaking of the PQ symmetry takes place before inflation. In general, this is a necessary cosmological requirement if the breaking of the original continuous symmetry leaves as a remnant a discrete symmetry $(Z_N)$ \cite{Sikivie:1982qv}, due to the formation of domain walls.

In the PQ case the axion is characterized by the Wess-Zumino (WZ) term
 \ba
\mathcal{L}_{anom} = \frac{g^2}{32 \pi^2}
\, \frac{a}{f_a} \, G \, \tilde{G}  \, ,
\ea
and there is a strict relation between the mass of this particle, $m_a$,  and the axion decay constant $f_a$,
\ba\label{rel-f}
m_a = \frac{z^{1/2}}{1+z} \frac{f_\pi \, m_\pi}{f_a}
= 0.6 \cdot 10^{-3} \,
\left(\frac{10^{10} \; {\rm GeV}}{f_a}\right)
\; {\rm eV} \, ,
\ea
where $f_\pi = 92$~MeV and $m_\pi = 135$~MeV are respectively the pion decay constant and mass, while $z= m_u/m_d \approx 0.6$ is the up to down quark mass ratio. Mass and coupling to gauge fields, for a global anomalous $U(1)$ theory, are related by the same scale $f_a$. In the case of anomalous $U(1)$ gauge currents this limitation is absent due to the presence of independent mass/coupling relations, a property that comes directly from the gauging of the axionic symmetry.

 Nambu-Goldstone modes of the CP-odd sector of some generic potential, generated because of the breaking of the original symmetry at a very large scale, as well as axions from string theory, share these descriptions, both inheriting specific
Wess-Zumino interactions in their effective actions, which are indeed the most distinctive features of their couplings to the gauge fields.
Generically, in a gauge theory, WZ terms are naturally generated by the decoupling of a heavy chiral fermion, due to the large vev of a Higgs, and leaving at lower energy an effective action which can be expanded in the mass
of the associated gauge boson (the gauge boson to which the fermion couples) or of the heavy fermion. In the case of abelian interactions the effective field theory takes a typical St\"uckelberg form \cite{Coriano:2007fw,Coriano:2007xg, Coriano:2009zh}. The presence of WZ terms guarantees the gauge invariance of the effective action at 1-loop - and indeed to all orders - since anomalies are just 1-loop effects. The theory has a unitarity bound \cite{Coriano:2008pg}. In this approach, the non-local structure of the anomaly is hidden in the infinite higher dimensional terms which are needed
for a consistent cancellation of the anomalies in a gauge theory. Phenomenological studies of models with dimension-5
$a F\tilde{ F}$ WZ terms, where $a$ is the axion, in the case of gauge anomalies, have been discussed both in non-supersymmetric \cite{Coriano:2005js,Coriano:2008wf} and in supersymmetric scenarios.
The identification of a physical axion in these types of models requires an extended superpotential, such as the one of the NMSSM, containing an extra singlet superfield, and can be realized by introducing an extra anomalous $U(1)$ gauge symmetry \cite{Coriano:2008xa}. In the MSSM, for instance, the axion takes the role of a goldstone mode \cite{Anastasopoulos:2008jt} and Higgs-axion mixing is absent. The mixing is necessary in order to generate a physical axion after electroweak symmetry breaking. This supersymmetric construction can be used in the study of models containing gauged axions and neutralinos as dark matter \cite{Coriano:2008xa}.

\section{Bottom-up approaches and anomaly poles}
In these types of effective actions the non-local structure of the anomaly is not immediately noticeable, nor is its connection with some specific massless degrees of freedom which appear quite clearly in the perturbative expansion. In fact, the completion of an anomalous theory, if not achieved by the addition of extra fermions which turn massless as
we move upward the cutoff of the effective theory, requires a non-local extension, since anomalies cannot be canceled by local counterterms 	\cite{Armillis:2008bg}.
One can ask, in this respect, simple questions and investigate the answer in models where the entire non-local nature of the anomalies appears, instead, quite visibly. This approach is rather different from the one mentioned in the previous section, being more radical, since in this case we do not allow an expansion scale \cite{Armillis:2009sm}. In this sense,
an anomalous correlator (and the associated effective action) represents a significant variant respect to the usual Wilsonian expansion, which is formulated in terms of local operators, because of the appearance of (massless) pole terms. The only possible way by which the pole term can be expanded is by a mass shift, as in the case of the pion pole which interpolates between the $U(1)_A$ QCD current and the vector currents, mediating its decay. The underlying mechanism involved is non-perturbative, but arguments based on anomaly matching reassure that the anomaly computed by using the pion pole (the new degrees of freedom of the effective theory) in the anomalous correlator and the perturbative anomaly coincide. Obviously, this does not indicate a complete overlap between two descriptions, since the anomaly loop carries no information about the structure of the QCD vacuum (power corrections, condensates, and so on).

Following these arguments, one may consider an anomalous theory in the low energy domain and ask for the structure of its completion at high energy, in the absence of intermediate scales (for instance in the form of new heavy fermions of mass $m$) as one moves upward the energy cutoff. It is then obvious that the non-local character of the anomaly, in this case, shows up to its fullest extent via the anomaly pole, and for this reason the usual $1/m$ expansion of the lagrangian is not appropriate, and should necessarily be replaced by an alternative expansion in terms of a non-local contribution (the pole term) plus $O(m)$ corrections \cite{Armillis:2009sm}. This is the variant to Wilson's approach mentioned above.

The absence of an expansion scale in theories which are ``just anomalous" at low energy and which ask for a completion in the UV, shows that the completion has necessarily to
erase the most direct signature of the anomalous low-energy action, contained in its anomaly poles\cite{Armillis:2009im}, which are, indeed, the common signatures of gauge, conformal, and probably also gravitational anomalies \cite{Giannotti:2008cv}. Perturbation theory and the unitarization of the effective action, as indicated by the pole structure, in this case, should not be taken as a replacement for the non-local completion, which is probably a string theory or else, and the mechanism of ``pole subtraction" is a hint about the presence of massless pseudoscalar degrees of freedom coming from the completion. These types of conclusions emerge from a perturbative analysis of the off-shell effective action \cite{Armillis:2009sm}.

While in the case of chiral gauge theories, the disappearance of the pole is somehow necessary for ensuring the unitarization of the effective theory at high energy, in the case of conformal anomalies the corresponding poles, which, recently, have also been found  \cite{Giannotti:2008cv, Armillis:2009pq} play a different role, since gravity breaks unitarity in the UV already at Born level, and there is no compeling need to relate the cancellation of these contributions to ensure unitarization of the theory .

The local formulation of these types of theories requires two additional scalar (for conformal) or two pseudoscalar (for gauge anomalies) degrees of freedom in order to rewrite these polar interactions in a local form, one of them being a ghost in both cases \cite{Coriano:2008pg}. In the case of a chiral gauge anomaly, this can be easily worked out resorting to a diagrammatic expansion, since the pole is ``unfolded" in terms of two St\"uckelberg axions $a$ and $b$, as discussed in \cite{Coriano:2008pg }
\beqa
\mathcal{L} &=& \overline{\psi} \left( i \slashed{\partial} + e \slashed{B} \gamma_5\right)\psi - \frac{1}{4} F_B^2 +
\frac{ e^3}{48 \pi^2 M} F_B\wedge F_B ( a + b) \nonumber \\
&& + \frac{1}{2}  \left( \partial_\mu b - M B_\mu\right)^2 -
\frac{1}{2} \left( \partial_\mu a - M B_\mu\right)^2.
\label{fedeq}
\eeqa
A detailed analysis shows that the functional integrals on $a$ and $b$ are gaussians and one recovers the pole contribution after functional integration.
Notice that $b$ has a positive kinetic term and $a$ is ghost-like. Both $a$ and $b$ shift by the same amount under a gauge transformation of $B$
\beq
a\rightarrow a + M \theta,\,\,\,\,\, b\rightarrow b + M \theta
\eeq
where $\theta$ is a gauge parameter. Similar conclusions are reached in \cite{Giannotti:2008cv}.

Both in the case of anomalous gauge theories (for instance in theories with anomalous $U(1)$'s) and in the coupling of the Standard Model  to gravity, where the presence of a conformal anomaly induces anomaly poles in the effective action due to the gauge-gauge-graviton vertex \cite{Giannotti:2008cv,Armillis:2009pq}, the introduction of auxiliary scalars and pseudoscalar fields may help in understanding the interplay between the problem of the stability of the vacuum and the coupling of these theories to extra sectors which may, eventually, remove the ghost from the spectrum.
The instability of the vacuum, in the case of anomalous gauge theories, are signalled by the ghost-like nature of one of the two axions, or of one of the two (extra) scalars introduced to render local the conformal anomaly. We conclude
that pole singularities have a special meaning and can be kept separate from the remaining corrections in the effective action.
 In this respect, the usual $1/m$ Euler-Heisenberg expansion of the QED effective action \cite{Armillis:2009im}, for an anomalous $U(1)$ theory, should be replaced by a different structure, in which we separate the pole contribution from the additional mass corrections.
The complete action, in this case, is instead given by
\beq
 \Gamma^{(3)}=  \Gamma^{(3)}_{pole} + \tilde{\Gamma}^{(3)}
 \eeq
 with the pole part given by
\beq
\Gamma^{(3)}_{pole}= -\frac{1}{8 \pi^2} \int d^4 x \, d^4 y  \,\partial \cdot B(x) \square^{-1}_{x,y} F(y) \wedge F(y)
\label{gammapole}
\eeq
and the rest ($\tilde{\Gamma}^{(3)}$) given by a complicated non-local expression which contributes homogeneously to the Ward identity of the anomaly graph. This second contribution can be expanded in terms of local operators. We omit technical details that can be found in \cite{Armillis:2009im}.

\section{Conformal case}
As we have already mentioned, in the case of the conformal anomaly, one discovers, analogously to the chiral case, the appearance of massless degrees of freedom \cite{Giannotti:2008cv} in the effective action. These appear after replacing the anomaly pole with local degrees of freedom. An off-shell analysis shows that these singularities are indeed present under any kinematical conditions, even though they can be decoupled
in the IR \cite{Armillis:2009pq} (with a zero residue). A discussion of their coupling to the effective action of gravity can be found in \cite{Giannotti:2008cv}.
The identification of anomaly poles in the $TJJ$ correlator under general kinematics, which are responsible for the appearance of a conformal anomaly, allows us to perform a separation between massless and massive contributions to the gravitational effective action  \cite{Armillis:2009pq, Giannotti:2008cv}. The extension of this analysis to the case of a chiral fermion will allow to derive the complete off-shell effective action which describes the coupling of the Standard Model to gravity mediated by the conformal anomaly.

\section{Conclusions}
There are several interesting points which still need to be elucidated in the case of anomalous gauge theories and which have not been fully grasped or understood, especially in their phenomenological implications. The ubiquitous presence of anomalous $U(1)$ symmetries in string compactifications,
either in top-down or in bottom-up formulations, is connected to massless scalar and pseudoscalar degrees of freedom which may acquire a potential non-perturbatively. On a similar line, conformal anomalies, which appear in the coupling of gravity to the Standard Model can be attributed to the appearance of anomaly poles and hence, by a local reduction, to two
scalar degrees of freedom. Massless scalars and pseudoscalars may play a significant role in the phenomenology of the early universe as well as in collider searches \cite{Coriano:2009zh,Armillis:2008vp}.

\begin{acknowledgement}
We thank G. Lazarides for discussions. This work is supported in part  by the European Union through the Marie Curie Research and Training Network ``Universenet'' (MRTN-CT-2006-035863).
\end{acknowledgement}
\providecommand{\WileyBibTextsc}{}
\let\textsc\WileyBibTextsc
\providecommand{\othercit}{}
\providecommand{\jr}[1]{#1}
\providecommand{\etal}{~et~al.}

\end{document}